\documentclass[review]{cas-sc}

\usepackage[numbers]{natbib}

\usepackage{longtable} 
\usepackage{changepage} 
\usepackage{multirow}

\usepackage[export]{adjustbox} 

\newcommand{\todo}[1]{}
\renewcommand{\todo}[1]{{\color{red} TODO: {#1}}}

\def\tsc#1{\csdef{#1}{\textsc{\lowercase{#1}}\xspace}}
\tsc{WGM}
\tsc{QE}
\tsc{EP}
\tsc{PMS}
\tsc{BEC}
\tsc{DE}

\begin{document}
\let\WriteBookmarks\relax
\def\floatpagepagefraction{1}
\def\textpagefraction{.001}

\shorttitle{RE Framework for Human-centered AI-based Software Systems}

\shortauthors{Ahmad et~al.}

\title [mode = title]{Requirements Engineering Framework for Human-centered Artificial Intelligence Software Systems}                      



\author[1]{Khlood Ahmad}


\ead{ahmadkhl@deakin.edu.au}

\affiliation[1]{organization={Deakin University}, 
    city={Geelong},
    state={VIC},
    country={Australia}}
    
\author[1]{Mohamed Abdelrazek}
\ead{mohamed.abdelrazek@deakin.edu.au}

\author[2]{Chetan Arora}
\ead{chetan.arora@monash.edu}

\affiliation[2]{organization={Monash University}, 
    city={Clayton},
    state={VIC},
    country={Australia}}

\author[1]{Arbind {Agrahari Baniya}}
\ead{aagraharibaniya@deakin.edu.au}

\author[3]{Muneera Bano}
\ead{muneera.bano@}

\affiliation[3]{organization={CSIRO's Data 61}, 
    city={Clayton},
    state={VIC},
    country={Australia}}

\author[2]{John Grundy}
\ead{john.grundy@monash.edu}

\begin{abstract}
 \noindent [Context] Artificial intelligence (AI) components used in building software solutions have substantially increased in recent years. However, many of these solutions focus on technical aspects and ignore critical human-centered aspects. [Objective] Including human-centered aspects during requirements engineering (RE) when building AI-based software can help achieve more responsible, unbiased, and inclusive AI-based software solutions. [Method] In this paper, we present a new framework developed based on human-centered AI guidelines and a user survey to aid in collecting requirements for human-centered AI-based software. We provide a catalog to elicit these requirements and a conceptual model to present them visually. [Results] The framework is applied to a case study to elicit and model requirements for enhancing the quality of 360\textdegree~videos intended for virtual reality (VR) users. [Conclusion] We found that our proposed approach helped the project team fully understand the human-centered needs of the project to deliver. Furthermore, the framework helped to understand what requirements need to be captured at the initial stages against later stages in the engineering process of AI-based software.
\end{abstract}



\begin{keywords}
Requirements Engineering \sep Software Engineering \sep Artificial Intelligence \sep Machine Learning \sep Human-centered \sep Conceptual Modeling \sep Virtual Reality \sep Empirical Software Engineering
\end{keywords}

\accepted{26 April 2023}

\maketitle

\section{Introduction}~\label{sec:introduction}
AI-based software systems are rapidly becoming essential in many organizations~\cite{amershi2019software}. However, the focus on the technical side of building artificial intelligence (AI)-based systems are most common, and many projects, more often than not, fail to address critical human aspects during the development phases~\cite{maguire2001methods,schmidt2020interactive}. These include but are not limited to age, gender, ethnicity, socio-economic status, education, language, culture, emotions, personality, and many others~\cite{grundy2021impact}. Ignoring human-centered aspects in AI-based software tends to produce biased and non-inclusive outcomes~\cite{amershi2014power}. Shneiderman~\cite{shneiderman2021human} emphasizes the dangers of autonomy-first design in AI and the hidden biases that follow. Misrepresenting human aspects in requirements for model selection and data used in training AI algorithms can lead to discriminatory decision procedures even if the underlying computational processes were unbiased~\cite{calders2013unbiased}. For example, a study by Carnegie Mellon revealed that women were far less likely to receive high-paying job ads from Google than men~\cite{lloyd2018bias} due to the under-representation of people of color and women in high paying IT jobs.

Studies on human-centered design aim to develop systems that put human needs and values at the center of software development and clearly understand the context of the software system's usage~\cite{maguire2001methods,shneiderman2022human}. These human values include security, tradition, achievement, power, etc.~\cite{schwartz2012overview}. An increasing number of research resources are being invested in developing human-centered AI solutions. Large organizations such as Google, Microsoft, and Apple are moving towards including human-centered values when building AI-based software~\cite{GooglePair2019, Microsoft2022, Apple2020}. These large organizations have now devised explicit guidelines for building human-centered AI systems. While these guidelines provide an excellent platform, they mainly focus on the design phase of software development and not on requirements engineering (RE).   

Recent research efforts in RE have focused on including human-centered requirements in building software solutions, such as emotions~\cite{kissoon2019emotion, ramos2005emotion, maiden2012spocks,thompson2013evoking, miller2012understanding, callele2008emotional},  gender~\cite{burnett2016finding, vorvoreanu2019rom}, power and politics~\cite{milne2012power}, personality traits~\cite{kanij2015empirical, soomro2016effect}, age~\cite{mcintosh2021evaluating}, and mental and physical challenges~\cite{grundy2018supporting}.   For example, Miller et al.~\cite{miller2015emotion} argued that ``emotions should be considered as first-class citizens in software engineering methodology'' and acknowledged the fact that software engineers usually overlook emotion when building software systems. Other studies have stressed that emotions are usually neglected when it comes to RE~\cite{colomo2010study}. Perera et al.~\cite{perera2020study} emphasized the importance of including human values early on in RE.

Although this is a growing research area in RE, limited studies focus on including requirements for human-centered AI-based systems. We examined the studies presented in our systematic literature review (SLR)~\cite{ahmad2021s, ahmad2022mapping} and a mapping study~\cite{villamizar2021requirements} on RE for AI (RE4AI) and found that most AI-based software lacks human-centered approaches when writing and modeling requirements, and existing research mainly focused on explainability, trust, and ethics with limited empirical evaluations. 

We propose a new framework, the RE for human-centered AI (RE4HCAI), to specify and model requirements for human-centered AI-based software. The RE4HCAI framework is based on industrial human-centered guidelines, an analysis of studies obtained from the literature, and the results from an experts' survey~\cite{ahmad2023requirements}. The survey identified the guidelines and human-centered aspects that should be addressed and prioritized during RE.
The framework is then evaluated in a case study, and requirements are elicited and modeled for a system that uses AI to enhance the quality of {360\textdegree} videos for virtual reality (VR) users. The purpose of the AI-based software was to improve the quality of experience (QoE) and quality of service (QoS) for systems streaming and rendering {360\textdegree} video contents. We discovered that some requirements could not be identified during the early stages of the project due to the black-box nature of the AI. The difficulty in explaining how it would respond to the available data also contributed to this issue. Therefore, a more iterative RE approach should be used to write requirements for AI-based~software.

\textcolor{Black}{The key contributions of this research can be described as follows:}
\begin{enumerate}
\item We present a new framework to help elicit and specify requirements for AI-based software. The framework is based on the industry's human-centered AI development~\cite{ahmad2022mapping}, an SLR~\cite{ahmad2021s}, and an expert survey~\cite{ahmad2023requirements}. As part of our framework, we provide a catalog to aid the elicitation of requirements for AI-based software. The framework further provides a conceptual model to present the requirements visually.
\item We apply and evaluate the framework on a case study that uses AI to enhance the spatial quality of {360\textdegree} VR videos and report on our findings.
\end{enumerate}

The rest of the paper is structured as follows:  Section~\ref{sec:Background}
provides a brief background related work. Section~\ref{sec:Framework} presents details of our framework. Section~\ref{sec:CaseStudy} reports on implementing the framework and modeling language on a software system to enhance {360\textdegree} videos for VR users. Section~\ref{sec:Discussion} discusses key results and summarizes emerging theories. Section~\ref{sec:Threats} addresses threats to validity. Section~\ref{sec:RelatedWork} discusses related work, and section~\ref{sec:Conclusion} concludes.

\section{Background and Motivation}~\label{sec:Background}

As mentioned in section~\ref{sec:introduction}, our framework is aimed at helping the elicitation and specification of requirements for human-centered AI-based software. We conducted an SLR \cite{ahmad2021s,ahmad2022mapping} and selected any existing study that focused on human-centered approaches in RE4AI. These papers include studies on emotion for a human-centered social robot~\cite{bruno2013functional}, understating expectations and limitations of AI-based software~\cite{sandkuhl2019putting}, identifying and mitigating human-centered issues related to fairness and biases~\cite{fagbola2019towards}, ethics~\cite{kuwajima2019adapting, aydemir2018roadmap}, explainability~\cite{hall2019systematic, schoonderwoerd2021human, cirqueira2020scenario, kohl2019explainability}, and trust~\cite{amaral2020ontology}.  Next, the industrial guidelines on human-centered AI development are analyzed, specifically Google's PAIR guidebook~\cite{GooglePair2019},  Apple's human interface guidelines for building ML applications~\cite{Apple2020}, Microsoft's eighteen guidelines for human-centered AI interaction~\cite{amershi2019guidelines} and the machine learning (ML) canvas~\cite{MLCanvas}. We note that ML canvas does not directly address the human-centered aspect. Nevertheless, we chose to include ML canvas for its relevance to our work as the ML canvas complemented the industrial guidelines by providing means for capturing relevant information for building ML models, e.g., ML business needs, decisions, data sources, and evaluation, and it facilitates collaboration between different stakeholders. 

The gathered requirements from the SLR and the industrial guidelines are summarized, mapped, and categorized into six areas to build human-centered AI-based software. 
The categorization of six areas aligns with the five categories proposed in Google PAIR. The sixth area, i.e., model needs, identifies the human-centered approaches used when selecting and training an appropriate AI-based~model~\cite{ahmad2023requirements}. Each area is explained in detail below as follows:

\subsection{Area\#1 Requirements for User Needs} 

The first area focuses on ensuring that the user needs are captured first when building AI-based software, documenting the limitations and capabilities, what the system can do, how well it can do its assigned tasks, and identifying the users and stakeholders. Other factors include how the user will interact with the system and what approaches should be selected to match the user's needs. Is the system going to be proactive or reactive? For example, in a proactive system, Google search provides a list of results based on the users' entered keywords. The next step is to determine if the user is aware of the AI feature. Visible features are when the user is aware of the AI component, such as having auto-complete trying to guess what the user wants to input next. Invisible features are when the user is not aware of the AI feature. For example, a map would gather real-time data on traffic to provide the best route to users~\cite{GooglePair2019,Microsoft2022,Apple2020, MLCanvas}. 

Next, we need to determine if the system is going to automate or augment the user's needs. For instance, AI-based software can automate any task that does not require human oversight. Tasks that a user enjoys or requires a human-in-the-loop can be augmented by AI to improve the user's experience and efficiency.

\textcolor{blue}{Finally, the evaluation behind the choice of reward function should be specified. How will the AI choose between right and wrong predictions of a learned model (LM)? An LM, as defined by Berry~\cite{berry2022requirements}, ``is the result of an instance of ML or deep learning (DL), whether the LM is taught, self-taught, or both with relevant real-world (RW) data''. In other words, an LM is a computational representation that is generated by an ML or DL algorithm. The model is learned by training on relevant RW data, which allows it to make predictions or decisions based on new, unseen data. The learning process typically involves feeding the algorithm a large dataset containing input-output pairs, from which the algorithm can discover patterns, relationships, or underlying structures in the data. During the training process, LM adjusts its internal parameters to minimize the error between its predictions and the actual outcomes. Once the model is trained and achieves a satisfactory level of performance, it can be applied to new, unseen data to make predictions, classify data points, or recommend actions, depending on the specific task it was designed to perform.}

\textcolor{blue}{In the case of binary classification, a prediction by an LM can result in four categories, i.e., true positives (TP), true negatives (TN), false positives (FP), and false negatives (FN). The two reward functions available are precision and recall.
Precision is the proportion of correctly picked TP out of the total positive predictions, i.e., precision = TP / (TP + FP). TP refers to the correctly predicted positive instances, while FP refers to instances where the LM predicts an incorrect positive outcome. A flight booking app with high precision would predict the cheapest flights for a specific date but might miss out on some flights. Recall measures the proportion of TP instances the LM can correctly identify. Thus, recall = TP / (TP + FN). Higher recall provides the confidence that the system has included all relevant results but might also include some irrelevant results \cite{GooglePair2019}.}
Evaluating the reward function will depend on the specific task and domain of the AI application used~\cite{berry2022requirements}. The decision to use either would be based on the trade-off between selecting precision and recall. Therefore, when evaluating the reward function, a list of trade-offs should be documented to justify the selected function.

\subsection{Area\#2 Requirements for Model Needs}

The second area includes documenting requirements for model needs. How do we choose an algorithm that optimizes stakeholder satisfaction? Do we need a system that is explainable or accurate? For example, Shin et al.~\cite{shin2019data} look for ways to reduce cost by experimenting with different algorithms and found that some algorithms could produce better outcomes with lower costs. However, these algorithms might lack in other aspects, such as explainability. In contrast, different algorithms might provide a better explanation but with predictions with lower confidence~\cite{krause2016interacting}. The choice of algorithm will also depend on the user's needs for the system. Also, as part of requirement specifications for AI, there is the need to specify the possible settings of each variable for a given AI algorithm when applied to a particular task in a given context~\cite{berry2022requirements}.

Part of the requirements specifications should consider how the incoming data affects model training. Thus, deciding how LM will improve on new data should be specified. Does the system need a dynamic LM that updates and trains online or a static system that improves only with updates? A dynamic system improves as the user interacts with the system and often involves using feedback in model training. A static system will improve with system updates and is trained offline. For example, an image recognition LM that depends on data that does not change over time will need updates with new releases, or else the LM might become biased or useless after a while~\cite{Apple2020}.

Model selection depends on the choice of algorithm and can include supervised, unsupervised, and reinforcement learning. When training the system, we need to specify a threshold to avoid overfitting or underfitting the data and select tools that will be used to evaluate the LM. When tuning the LM, the following should be considered: types of feedback, user behavior, training data used in model tuning, and adjusting the parameters accordingly \cite{Apple2020, MLCanvas, GooglePair2019}.

\subsection{Area\#3 Requirements for Data Needs} 

The third area focuses on data needs and data collection methods. Data collection includes the type and amount of data needed. Once data collection methods are specified, data requirements should be considered, including quality of data used, security, privacy settings, and fairness. Data quality includes five components: accuracy, completeness, consistency, credibility, and currentness. 1) Having correct data will ensure accuracy. 2) Completeness refers to the availability of all attributes and events associated with the data. 3) Consistency has no contradictions in the data used. 4) Ensuring credibility by having truthful data. 5) Finally, currentness means data must be collected within the correct time frame \cite{challa2020faulty}. 

Data quantity focuses on data being diverse. And data quality addresses the completeness, consistency, and correctness of data being used \cite{vogelsang2019requirements}. Selecting data should include identifying features, labels, and sampling rates. Labels identify the features needed to train the ML model, such as labeling a scanned image of a tumor as malignant. Explicit labeling is done manually, and implicit labeling is when the model learns the pattern independently. Examples represent a row of data and contain features, and labels represent descriptions given to data. More samples in the dataset ensure diversity but also increase costs \cite{shin2019data}. In this case, a threshold should be identified to set the amount of data needed within the given budget.   

Identifying and reporting biases in data must be addressed. Such biases can include automation, selection, group attribution, etc. Automation biases are when preferences are selected based on automated suggestions from the system. Selection bias usually happens when data is not collected randomly from the target population but rather selected based on the stakeholder's requests. Group attribution assumes that an output suitable for an individual will have the same impact on everyone in the group. Identifying key data characteristics should be set early to avoid discrimination and biases \cite{GooglePair2019, amershi2019guidelines}.
 
\subsection{Area\#4 Requirements for Feedback and User Control}

This area deals with identifying and finding which kinds of user feedback need to be established in RE when building AI-based systems. Different types of feedback include implicit, explicit, and calibration. Implicit feedback provides information about the user's interaction, preferences, and system behavior. Examples of implicit feedback include accepting or rejecting a recommendation, times of use, number of hours, when or how many times the user logged on, etc. The user provides explicit feedback when requested by the system, including surveys, forms, ratings, written feedback, likes, or dislikes. Calibration is the initial information the system might need from the user to function, such as scanning your fingerprint for the first time to activate touch~ID.

It is essential to identify how and when the feedback will be used in model tuning and what changes it will have on the AI-based system. When asking for feedback, privacy measures should be considered to secure it. Also, give the user a choice to dismiss the feedback. Allowing users to feel that they control the system is essential to human-centered AI and can depend on many factors. Providing the users with control can be achieved by giving them complete control of the system in case of a failure or providing them with multiple options from which they can make the appropriate choice~\cite{GooglePair2019, amershi2019guidelines, Apple2020}.

\subsection{Area\#5 Requirements for Explainability and Trust}

AI-based systems reasoning can include explaining the results or predictions to the end user and stakeholders. From the stakeholders' point of view, explainability requirements should consist of setting expectations of what the system can do and how it can do it and explaining its limitations and capabilities. On the other hand, explaining to the user will involve the data used and informing them about changes to the system that might happen with updates or LM improvements. Explaining predictions can be presented by either providing an example or displaying confidence. Confidence could be displayed in many ways depending on the situation and the AI-based system. Each situation should be evaluated to find an appropriate method to show confidence. 

Explainable requirements can often conflict with others, such as performance and cost. It might be cheaper to build systems that are not very explainable with better performance~\cite{kohl2019explainability}. Therefore, it is vital to calculate the trade-off when favoring an explainable system and identify how it might conflict with other requirements. Schoonderwoerd et al.,~\cite{schoonderwoerd2021human} indicate that explanations should be provided based on a specific context, with the need to identify which explanations should be provided and when. However, explainability might be necessary for some situations to ensure compliance with the European Union's General Data Protection Regulation (GDPR) \cite{goodman2017european, wachter2017counterfactual}. In this case, other measures need to be considered.

Providing realistic expectations helps users and stakeholders avoid over-trusting AI-based software. Explainability requirements might occur during the building process or after the LM is deployed. Explainability requirements can be divided into four components to include: Who the explanation is addressed to, what needs to be explained, when should the explanation happen, and who explains \cite{balasubramaniam2022transparency}.     Also, explanations should consist of consequences that might occur due to an action performed by the user. Showing the confidence can be a way to provide an explanation to users. However, it is essential to determine when and how to display predictions, as sometimes showing confidence could lead to mistrust \cite{GooglePair2019}.  

\subsection{Area\#6 Requirements for Errors and Failure}

None of the studies in the RE4AI literature mention the need to address errors in RE when building AI-based systems~\cite{ahmad2022mapping}. The focus of most industrial guidelines is on designing AI-based software with a human-centered approach and does not focus on RE. Our survey results~\cite{ahmad2023requirements} show that practitioners working on AI-based systems wanted to know how to deal with errors and specify error sources during RE. The different types of errors include background, context, and system limitations. Some errors, such as background and context, are more challenging to identify and are invisible to the end user. Context errors happen for several reasons and can be avoided by ensuring the user understands how the system works and ensuring the system is aligned with their needs. They are usually an outcome that is a true positive yet does not provide a prediction that aligns with the user's needs. 

Error sources include system, incorrect predictions, data, and input and output errors. Data errors happen with mislabeled data and can be due to poor training or inaccurate labeling. Prediction errors can occur when an incorrect model is used, the data is not comprehensive, or missing some critical elements. Input errors happen when users input unexpected data and can be due to users' old habits or having an abusive user. Output errors are when the system provides a prediction that is low in confidence or an irrelevant output but high in confidence. Moreover, finally, system errors would happen when multiple systems using AI  integrate or depend on each other. We need to identify error types and sources in RE and mitigate them by providing an action plan for how they should be addressed and fixed~\cite{Apple2020, GooglePair2019, amershi2019guidelines}.

\section{A Framework to Manage RE for Human-centered AI-based systems (RE4HCAI)}~\label{sec:Framework}

\begin{figure}[h!]
   \centering
\includegraphics[width=\linewidth]{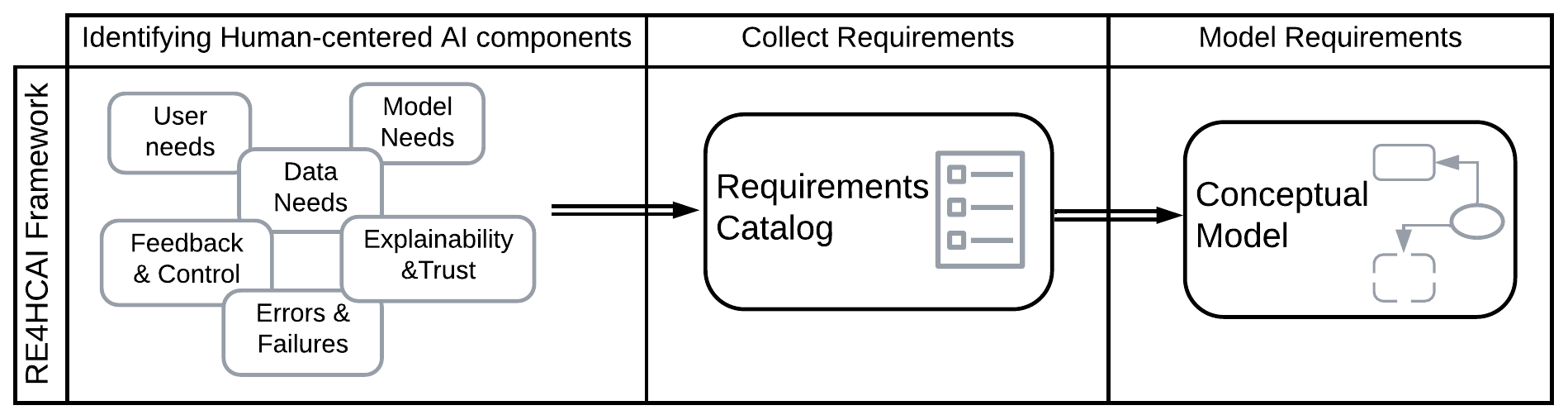}
\caption{Our proposed framework to elicit and model requirements for human-centered AI.}
\label{fig:RE4HCIFramework}
\end{figure}

This section presents the RE4HCAI framework, illustrated in figure~\ref{fig:RE4HCIFramework}, for eliciting and modeling requirements for human-centered AI-based systems. The framework consists of three layers. The first layer in section~\ref{sec:ReferenceMap} presents the human-centered guidelines that should be included in RE. The guidelines are combined and mapped into a reference model. The second layer presents a catalog (provided in the appendix) to help elicit requirements for building human-centered AI-based software and is discussed in section~\ref{sec:Catalog}. The last layer proposes a modeling language to present the requirements visually, as explained in section~\ref{sec:ModelingLanguage}.

\subsection{Human-centered AI Guidelines for RE}~\label{sec:ReferenceMap}

\begin{figure}[h!]
   \centering
\includegraphics[width=\linewidth]{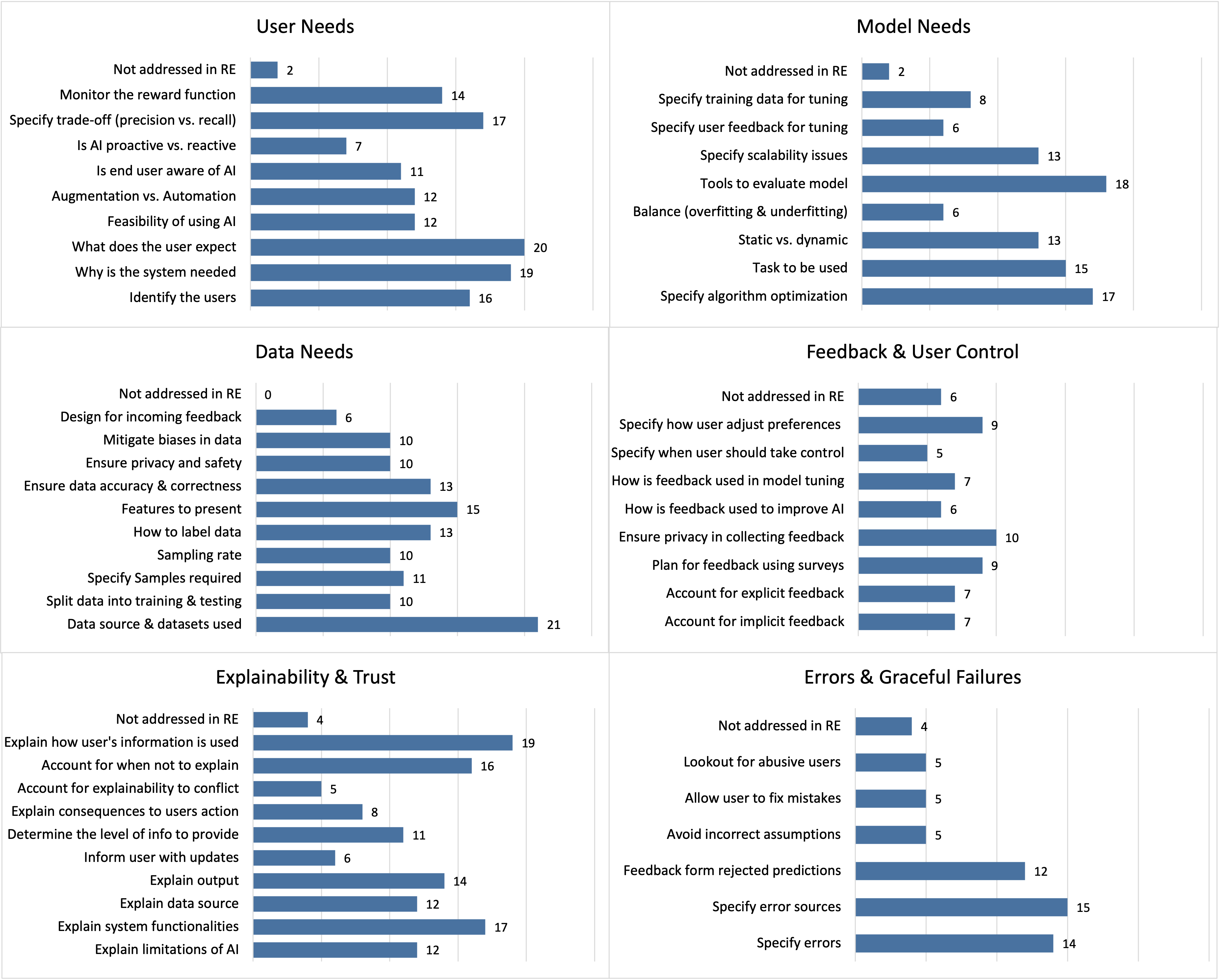}
\caption{Human-centered needs for AI and the frequency of survey participants who selected them.}
\label{fig:Needs}
\end{figure}

The coverage of these six areas in RE was investigated by conducting a user survey with 29 practitioners and researchers working on AI-related projects \cite{ahmad2023requirements}. The participants included data scientists, ML specialists, and software engineers working with AI-based software. We asked each participant to identify which of the mapped human-centered guidelines are or should be, included in RE based on their experience. The survey results, shown in figure~\ref{fig:Needs}, confirmed that all six human-centered areas need to be specified in RE4HCAI. The survey participants responded more favorably towards including the first three areas of user, model, and data needs than the remaining three areas for RE4HCAI. Thus, our evaluation focuses on these three areas in section~\ref{sec:CaseStudy}.

\begin{figure}[h!]
   \centering
\includegraphics[width=\linewidth]{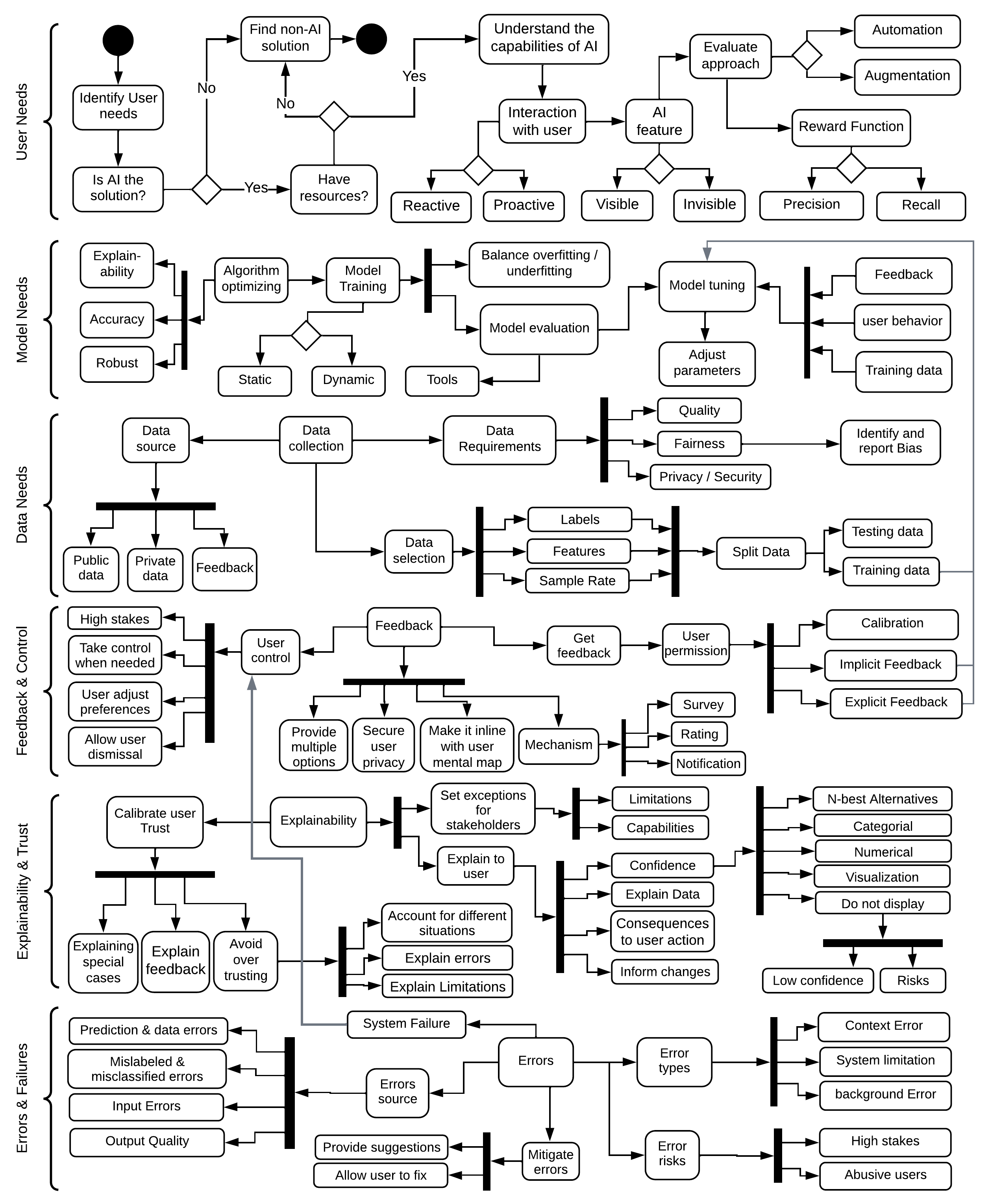}
\caption{Diagram showcasing the reference model for the RE4HCAI framework.}
\label{fig:Framework}
\end{figure}

The six areas of the human-centered AI components that are needed to be considered in RE are mapped to present a reference model that showcases the overall layout of the framework, as shown in figure~\ref{fig:Framework}. Each area presented in the reference model is explained further in section~\ref{sec:Background}.
 
\subsection{Catalog for Collecting Requirements}~\label{sec:Catalog} 
This section presents a catalog of human-centered AI requirements along the six areas in our framework. The gathered human-centered requirements and the mapped requirements from the SLR are listed in a tabular format to form our catalog. The catalog can be used as a checklist to elicit requirements for AI-based software and has six sections. Each section is dedicated to eliciting detailed requirements for each of the six areas explained in the background section~\ref{sec:Background}. The catalog is provided in the appendix.

\subsection{Modeling Requirements for Human-centered AI}~\label{sec:ModelingLanguage}

\begin{figure}[h!]
   \centering
\includegraphics[width=0.7\linewidth]{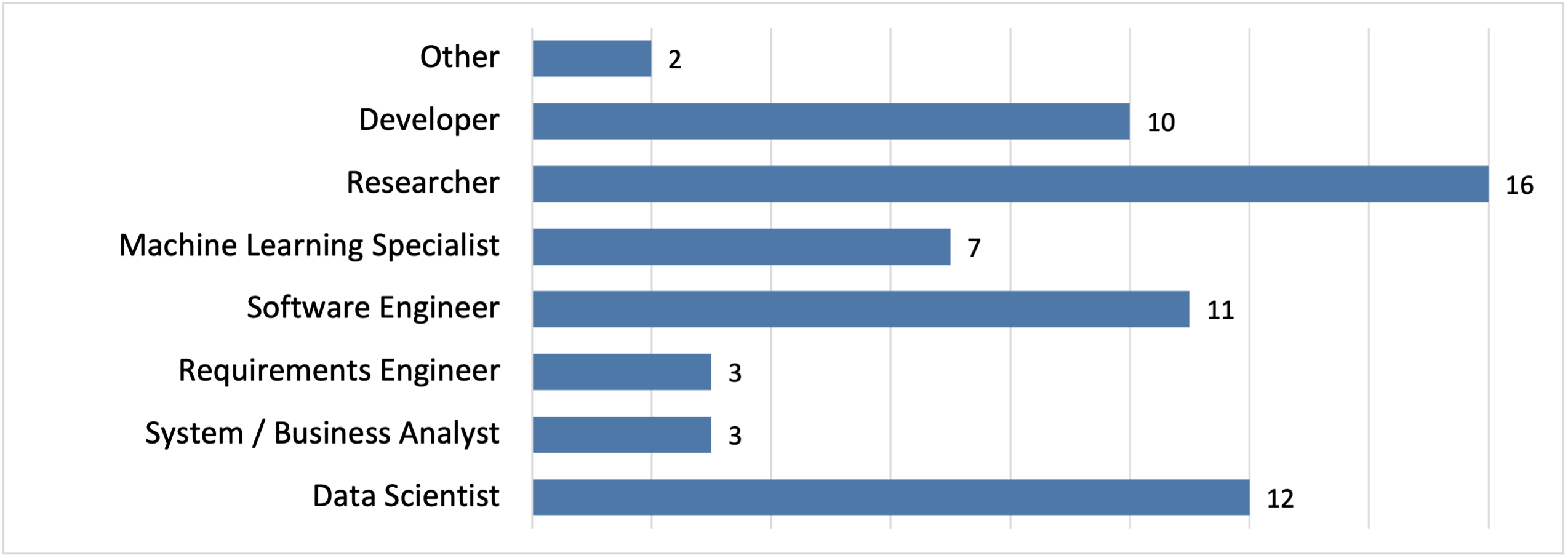}
\caption{The different roles of participants in the user survey.}
\label{fig:Roles}
\end{figure}

\begin{figure}[h!]
   \centering
\includegraphics[width=\linewidth]{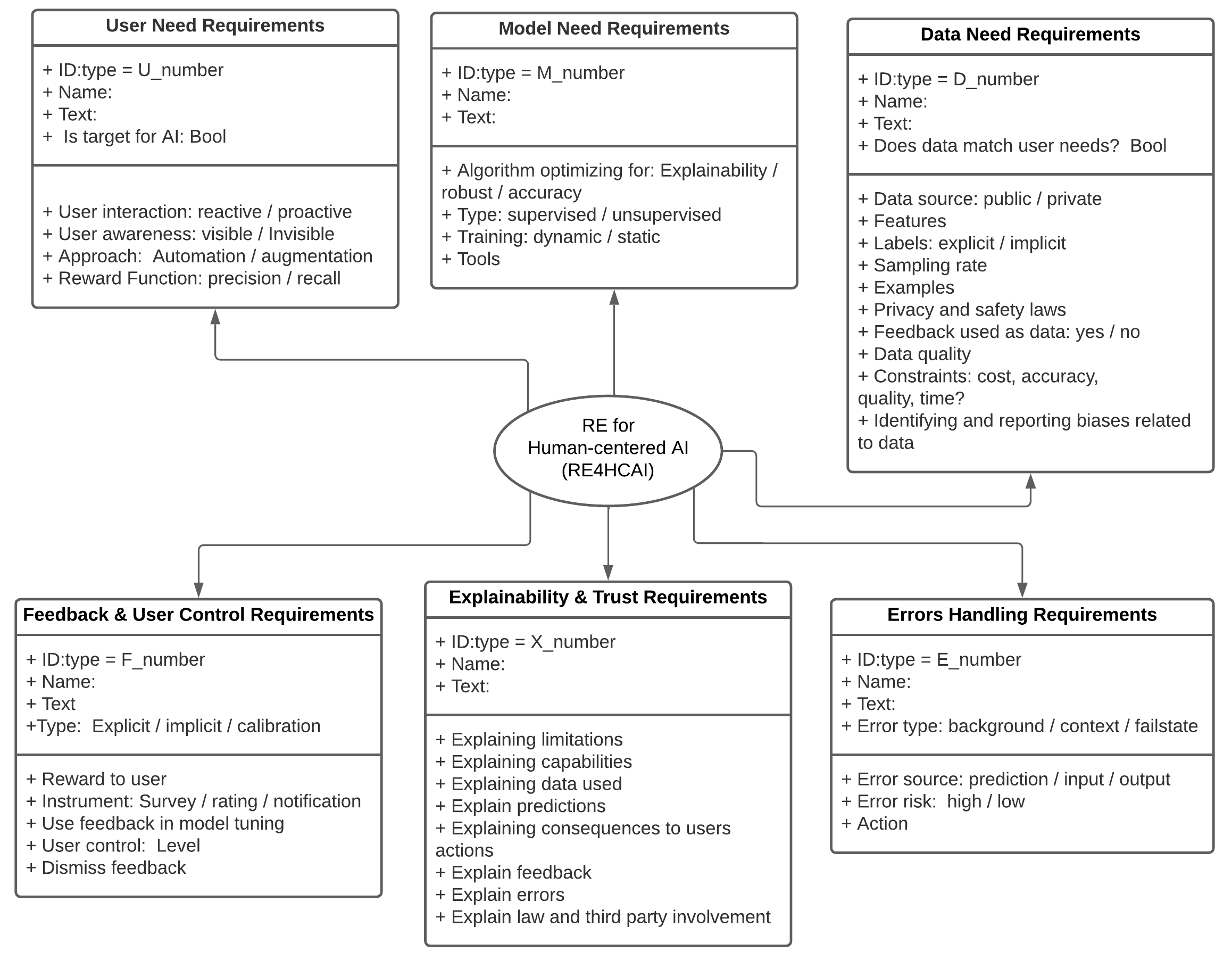}
\caption{First level of the RE4HCAI framework showing the six different areas.}
\label{fig:MetaModel}
\end{figure}

In our SLR, we sought any existing modeling language or requirements notations used in RE4AI. UML was the most popular method, making it easier for non-software engineers to work with and learn. However, UML has limitations as it does not support modeling non-functional requirements (NFR) and business rules. Goal-oriented requirements engineering (GORE) had better support for NFR and business rules but was more challenging to learn and was mainly used by requirements engineers and software engineers. Also, GORE could model requirements at lower levels of abstraction than UML. Silva et al.~\cite{silva2018requirements} explained that using GORE to model a requirement or concept could be presented with fewer structural diagrams than UML. However, UML is more widely known and used than GORE. Also, GORE was reported to be more challenging to learn among non-software engineers~\cite{neace2018goal,dimitrakopoulos2019alpha}.

Looking at the different roles from the survey, we found that around 25\% of the people involved in building AI-based software were software engineers and requirements engineers. Data scientists and ML specialists contributed around 27\% of the team building AI-based software. Other roles included system and business analysts, developers, and researchers, as shown in figure~\ref{fig:Roles}. Due to the diverse nature of team structure in building AI-based software and the need to use more specific modeling concepts to reflect on the process provided in the framework, we decided to create a conceptual model for the RE4HCAI framework, as conceptual modeling allows for precise concepts to be modeled as well as presenting a holistic view of the application~\cite{embley2012handbook}.

\begin{table}[]
\caption{Legend}
\label{table:legend}
\label{}
\begin{tabular}{ p{2.5cm}  p{4cm} p{1cm} p{2.5cm}  p{4cm}  } \hline
\textbf{Notation} & \textbf{Explanation} & & \textbf{Notation} & \textbf{Explanation} \\ \hline
\\
 \includegraphics[width=0.15\textwidth, valign=t]{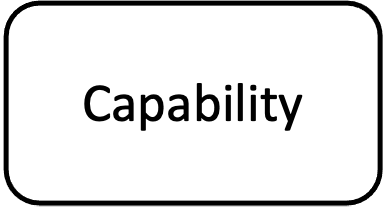}  & Displays a capability of the AI-based system (What the AI component of the system CAN do)  & &  \includegraphics[width=0.15\textwidth, valign=t]{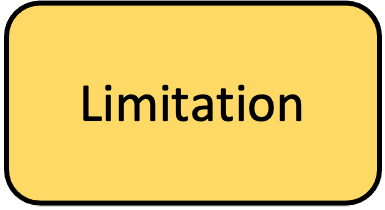} 
& Displays a limitation of the AI-based system (What the AI component of the system can NOT do)  \\ \\
\hspace{1.5em}\includegraphics[width=0.1\textwidth, valign=t]{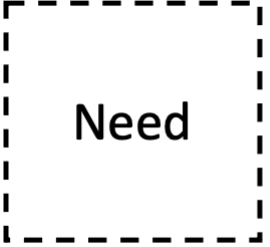}   & Display the need for the system & & \hspace{1em}\includegraphics[width=0.12\textwidth, valign=t]{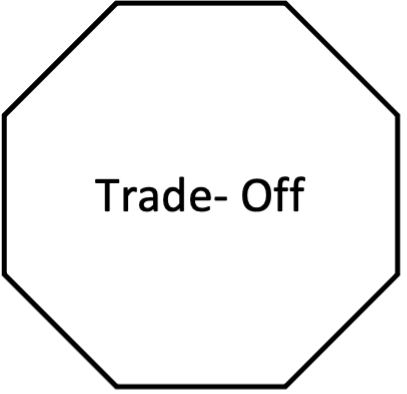}    & Showing the trade-off between two choices made (e.g., selecting precision vs. recall)\\ 
\includegraphics[width=0.15\textwidth, valign=c]{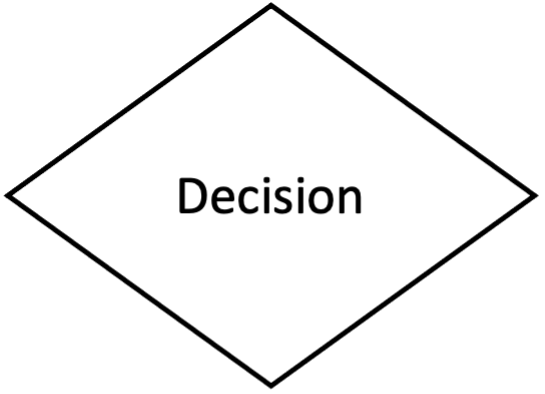} & Display a decision or choice to be made && \includegraphics[width=0.15\textwidth, valign=c]{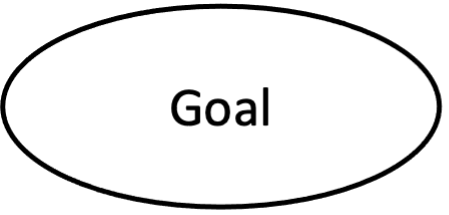} & Display the goals needed for the system to be built, goals can be functional requirements \\ 
\includegraphics[width=0.15\textwidth, valign=c]{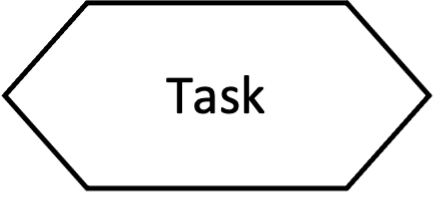} & A process or task that needs to be achieved &&    \includegraphics[width=0.15\textwidth, valign=c]{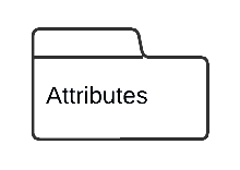}  & An attribute or property that is needed for a process \\
\includegraphics[width=0.15\textwidth, valign=c]{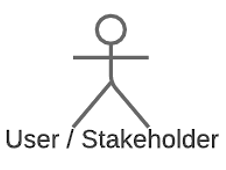}   & Shows the different users involved in building the systems or end-users & & \hspace{1em} \includegraphics[width=0.1\textwidth, valign=c]{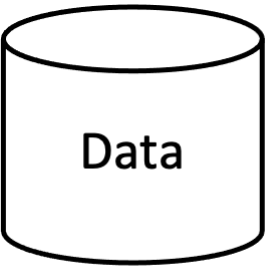} & Displays databases or data sources used in building the system \\ \\
 \includegraphics[width=0.15\textwidth, valign=t]{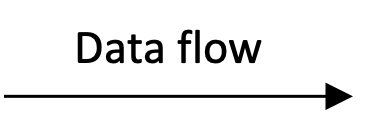}   & Direction the data flows in && \includegraphics[width=0.15\textwidth, valign=t]{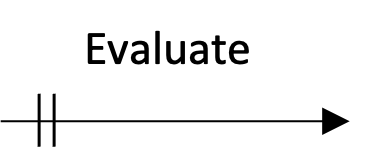}    & Data flow that needs to be evaluated \\ \hline
\end{tabular}
\end{table}

Our conceptual model consists of two levels. The first level presented in figure~\ref{fig:MetaModel} provides a holistic view of the system requirements.   We use an oval shape notation to list the main goal that needs to be addressed. For example, in the case study we conduct in section~\ref{sec:CaseStudy}, the goal would be enhancing the quality of 360\textdegree videos by four times. This goal would connect to the six sub-models to include the six areas of our framework, areas\#1-6 covered in section~\ref{sec:Background}. For each area, we use a UML class diagram to present it. Each area contains the main attributes in the reference model in figure~\ref{fig:Framework}. The second level consists of a sub-model to show each area in further detail. We provide the notations to model these requirements in table~\ref{table:legend}. When building the visual notations for our language, we attempt to comply with the nine principles of notations~\cite{moody2009physics} to reduce the complexity and cognitive load for visual notations in SE modeling. We present our notations using different shapes, textures, icons, and colors.

Goals are presented using an oval shape; each need includes the components modeled in the sub-models. Since our framework focuses mostly on needs, we use a square with dash lines to visually emphasize the concept of needs. Elongated hexagons are used to present processes or tasks. Trade-offs are displayed using an equilateral octagon. Limitations and capabilities are modeled using rectangles to show what the AI component can and cannot do, with limitations shaded in yellow to differentiate between them visually. Icons were used to present some notations.  The folder icon is used for attributes, the stick icon for users, and the data icon for data sources.  Moreover, finally, the components are joined using connectors to either show the flow of data or evaluate the data. 

\begin{figure}[h!]
   \centering
\includegraphics[width=0.5\linewidth]{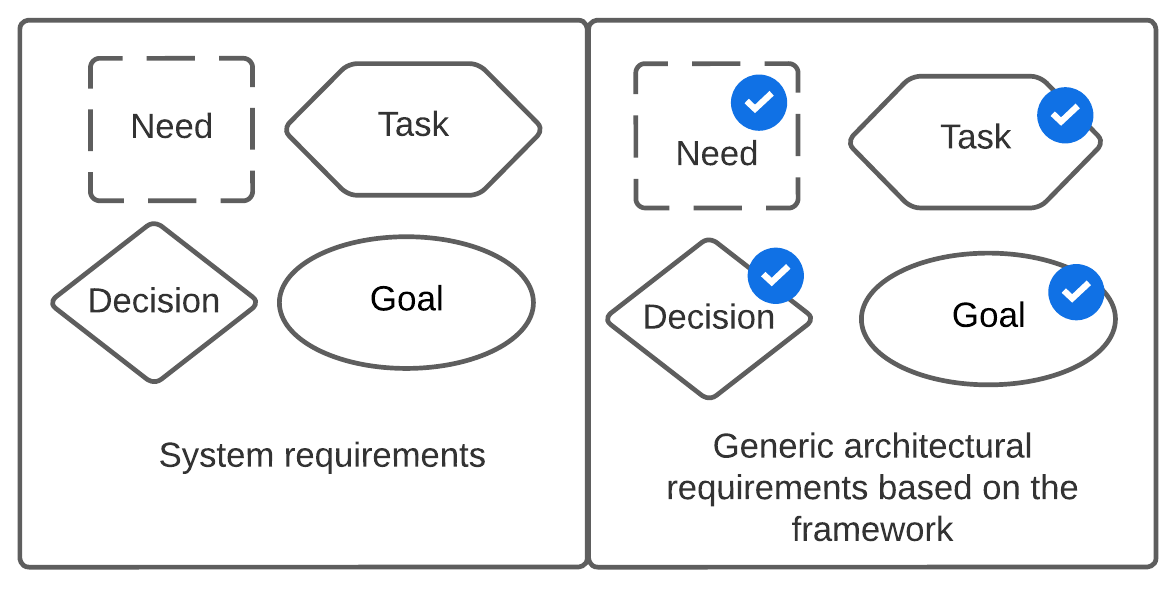}
\caption{The blue icon used with modeling notations shows the difference between architectural and system requirements.}
\label{fig:sysVSarch}
\end{figure}

When presenting the models, we use blue tick icons on notations to represent the parts that belong to the framework. The notations that do not include a blue tick icon are system specific, as shown in figure~\ref{fig:sysVSarch}. System requirements will change when modeling a different AI-based system.

\section{Case Study Application and Evaluation}~\label{sec:CaseStudy}

This section describes the process we used to design, select, and conduct the case study to implement our proposed framework. The case study answers the following research questions (RQs): 

\begin{enumerate}[RQ1:]
    \item How does our RE4HCAI framework benefit the process of eliciting and modeling requirements for human-centered AI software? RQ1 aims to assess the usefulness of our framework in eliciting and modeling requirements. It further contrasts the RE process before and after implementing our framework in the case~study.
    \item How can the RE practices be aligned in the life cycle of AI projects? RQ2 aims to provide an understanding and alignment of the RE practices in AI-based software development projects.
\end{enumerate}

The case study follows the guidelines for conducting case study research~\cite{kitchenham1995case} and empirical evaluations in software engineering~\cite{easterbrook2008selecting}. The main reason for conducting this study was to find how to extract requirements for building human-centered AI-based software and if implementing the framework benefited the process. The catalog is used to elicit requirements for the AI system, to identify user needs, model needs, and data needs as presented in tables~\ref{table:UserNeeds},~\ref{table:ModelNeeds},~and~\ref{table:DataNeeds}. These requirements are then modeled, as shown in figures \ref{fig:UserNeeds}, \ref{fig:ModelNeeds}, and \ref{fig:DataNeeds}.

\subsection{Study selection}

When selecting the case studies, we had the following selection criteria in mind: 
\begin{enumerate}

    \item Software project with a significant AI component: Since our framework focuses on RE4AI, we required a project with an AI component to answer our RQs, and, ideally, a project that involved interaction with users. 

    \item Project maturity: The project should be at a reasonably mature stage of development but ideally still under development. Our previous experience motivated this criterion, as we found that experts tend to miss out on details in already completed projects. Furthermore, to address RQ2, we required a project in which the requirements had been established using traditional methods.

    \item Availability and background of the expert: We needed to select a project with experts that had the requisite knowledge of the AI component of the system so we could elicit and model the requirements for model and data needs. Also, the case study required substantial time commitment from the expert(s), and hence we sought projects in which at least one expert agreed to the time commitment.    
    
\end{enumerate}

Our case study on VR-{360\textdegree} video enhancer matched the abovementioned criteria. The VR project builds on a DL model to enhance the quality of {360\textdegree} videos for VR platforms. Hence, it has a major AI component that satisfies the first criteria. Also, the project meets the second selection criteria as it was at the later stages of development. The project's requirements were finalized as a software feature plan, and most of the features were implemented when we conducted the case study. Furthermore, we had access to the primary ML expert of the project, who developed the AI component and agreed on the time commitment, which satisfies the third selection criteria.

\subsection{Data Collection}

We collected data over four sessions with the ML expert. Of these four sessions, three were dedicated to extracting and modeling the requirements for the project, and a fourth session evaluated the modeled human-centered requirements. The ML expert who built the AI project, and ended up being the fourth author, had no prior background knowledge of how our framework or the modeling language worked. They were given a brief overview of our framework in the first session. During the fourth session, we identified which requirements could be established at the start of the project. In other words, these were the requirements that the ML expert found important for them to know at the start of the project's lifecycle. However, some of these requirements were not necessarily known originally by the experts and were explored through our framework.

We note that we could elicit and model the requirements only for the first three areas of the framework collaboratively. This was the case due to the limited availability of the ML expert and the fact that the elicitation and modeling for each area required significant time commitment. While we ideally wanted to cover all six areas, we had a choice of covering a subset thoroughly or covering all areas superficially. We made the former choice, which aligns with our survey findings that practitioners' first three areas are the most focused areas when eliciting requirements~\cite{ahmad2023requirements}.

\begin{figure}[h!]
   \centering
\includegraphics[width=\linewidth]{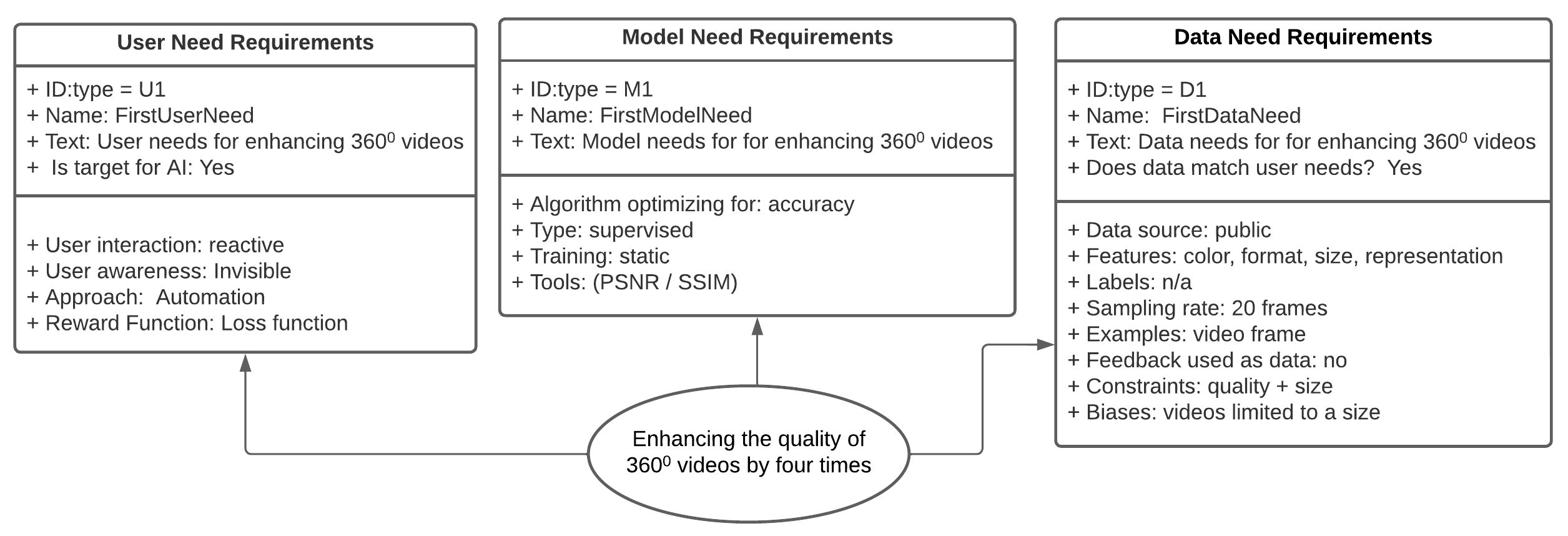}
\caption{Model presenting the holistic view of the requirements elicited for enhancing the quality of 360\textdegree video.}
\label{fig:Level1}
\end{figure}

In the first session, the ML expert gave an overview of the project, and the expert was given a brief overview of the motivation of the study. The expert was further given a brief overview of the framework. The first session was dedicated to collaboratively eliciting and then modeling the first area for user needs. The first session was the longest and lasted $\approx$4 hours and was divided into two periods. In the first half, the requirements were elicited, and the second half consisted of modeling the requirements. The second session was dedicated to the second area model needs. The session took $\approx$2.5 hours and was done over two periods. Similar to the first session, the requirements were elicited in the first half, followed by modeling the requirements in the second half. In line with the first two sessions, the third session was dedicated to eliciting and modeling the third area data needs. The session took $\approx$3 hours over two periods. The fourth session took two hours, and we evaluated all three models. We instructed the ML expert that they were free to suggest any changes to the elicited and modeled requirements from the three sessions. Also, the ML expert reflected on the comparison between the RE process followed originally in the project and the requirements collected from the sessions after applying our framework. The holistic view of the system is presented in figure~\ref{fig:Level1}.

 \subsection{Improving the Quality of {360\textdegree} Videos}

The case study was conducted on a project that uses DL to enhance the quality of {360\textdegree} videos for VR platforms \cite{agrahari2022omnidirectional}. Current approaches to building {360\textdegree} videos can capture and facilitate the immersive and interactive viewing experience. However, the quality of the final product can degrade due to the limitations of consumer-grade hardware, bandwidth streaming constraints, and processing that requires stitching videos from multiple cameras \cite{fan2019survey, anwar2020subjective, roberto2019visual}. The proposed solution was to enhance the quality of the final product using AI-based~software.

\subsection{Requirements for User Needs}

The first session was held to elicit requirements for identifying the user's needs. The session started by identifying the need for the proposed AI-based software using the catalog in the appendix. The need for the system was determined by the quality of the existing {360\textdegree} videos. Most {360\textdegree} videos had low resolution and contained artifacts and noise, thus affecting the quality of the final product \cite{huang20176, roberto2019visual}. The process of removing unwanted noise and artifacts was not achievable without the AI component, therefore, establishing the need for an AI-based software.

\begin{table}[]
\caption{Identified user needs requirements for the 360  video enhancer.  * requirements that could be specified before testing the data on the selected model}
\label{table:UserNeeds}
\centering
\footnotesize
\begin{tabular}{p{3cm} p{5cm} p{7cm}}
\hline
\multirow{3}{*}{Identify need for AI?} & Who are the users & * VR users who could watch enhanced 360 videos offline \\
& Why do we need the system &  * Improving the immersive experience for viewers\\
& What is the system used for & * Entertainment and teaching \\ \hline
\multirow{4}{*}{Systems capabilities} & \multirow{2}{*}{Limitations} & * Limitations to the user: the need for VR equipment with high rendering capabilities \\
 & & Limitations to stakeholders: Hardware resources and processing time \\
 & Capabilities & \begin{tabular}[c]{@{}l@{}} Improved quality of experience \\ 
 Improved quality of service \end{tabular} \\
 & How well can the system do what it does? & * It should improve the resolution of the video by four times \\
 \hline
Interaction with user action & Proactive (User requests action) or Reactive (Interacts with the user without requesting) & * Reactive – as it will refine the quality of the video without the users having to ask  \\
\hline
Is the user aware of the AI feature? & Visible or Invisible features & * Invisible feature, the user will not be aware of the AI component \\
\hline
Evaluate approach & Augmentation  vs. Automation  & * Automation – not possible for a human to be directly involved in super-resolving videos \\
\hline
\multirow{4}{*}{Reward function} 
& * A loss function & optimise model to predict correct pixel values\\
 & List potential pitfalls & Loss of visual quality \\
 & How do you provide inclusion & N/A \\ \hline
\end{tabular}
\end{table}

\begin{figure}[h!]
   \centering
\includegraphics[width=\linewidth]{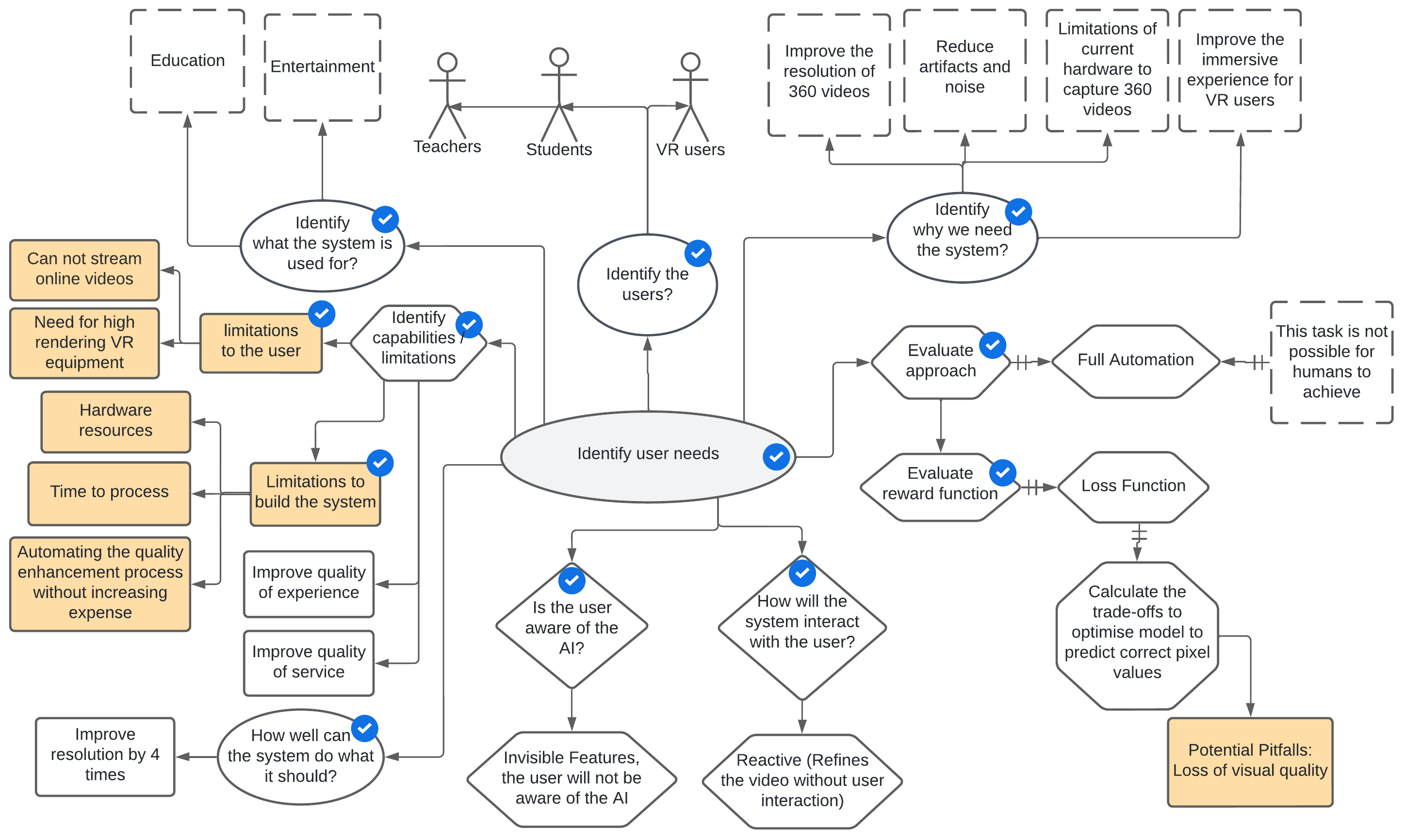}
\caption{Model presenting the requirements for User Needs.}
\label{fig:UserNeeds}
\end{figure}

To identify the system's capabilities, we needed to understand what the expected system could and could not do. System capabilities covered both the end-user and stakeholders. From the user's perspective, the end product would improve the resolution four times while using the same hardware equipment. Therefore, improving the QoE in either entertainment or educational settings. The limitations affected both the end-user and the stakeholders. Listing these limitations to avoid high expectations from either side was important. The limitations for the end-user were that users had to own VR equipment capable of rendering high-quality {360\textdegree} videos. The second limitation to the user was that they could watch these videos only offline. Therefore, online streaming would not be possible. Furthermore, the limitation for the stakeholder was towards building the system, which included hardware resources and time needed to process the videos. The stakeholders needed to establish ways to improve quality without increasing expenses, which will later link to model selection and collected data.   

When modeling limitations and challenges, we found that using color helped distinguish between them visually, as shown in figure~\ref{fig:UserNeeds}. We modeled the human-centered aspects as goals, and each goal would be mapped to either a need, process, capability, or limitation. For example, the system needed to be automated because it was not possible for a human to do the system's task. Processes involved actions that needed to be implemented, such as building a reactive system that doesn't require user interaction. Also, having different textures for the needs made it easier to identify them visually. The advantage of modeling the first area was that it was easy to recognize limitations, capabilities, and needs when building human-centered AI-based software.

\subsection{Requirements for Model Needs}

\textcolor{blue}{The second session involved collecting requirements for model needs from our catalog, which helped us understand the rationale behind selecting the type of algorithm used to train the LM and other aspects considered during optimization. The LM was built using deep learning techniques and as a regression problem and optimized for accuracy in order to produce better-quality frames.} 
Model tuning included adjusting the LM to user feedback, user behavior, and output errors. For this particular case, user feedback would be done with a ranking scale that is given after the user watches the enhanced video. The other form of feedback would be to perform a user study to evaluate the quality of the generated outcome from the model. The study would involve human subjects ranking the quality on a scale of 1 to 100. This feedback will indicate if the human-centered perceptual quality of the predicted outcome is acceptable. If not, the plan would be to change the model training and optimize it to improvise perceptual quality. We note that we modeled as limitations the cases where user or system goals could not be achieved, e.g., tracing LM output errors back to data.

\begin{table}[]
\caption{Identified model needs requirements for the 360 video enhancer.  * Requirements that could be specified before testing the data on the selected model}
\label{table:ModelNeeds}
\centering
\footnotesize
\begin{tabular}{p{3cm} p{4cm} p{8cm}}
\hline
Optimize for? & Accuracy & * Compare the ground truth of data to the predictions \\ \hline
Choose the ML type & Supervised  &* Regression  \\ \hline
Model training & Static with offline training & * Fixed dataset that the model learns from\\ \hline
Balance between overfitting \& underfitting & Keep track of training losses & Pinpoint the iteration at which the model when overfitting starts  \\ \hline
\multirow{3}{10em}{Model Tuning (Include parameter tuning and architecture changes)}  & User feedback & Provide a ranking to evaluate the perceptual quality \\
 & Adjusting to user behavior & Change model training to optimize losses based on results from a user study to evaluate perceptual quality \\
 & Trace output errors to data & In case of output errors occur, data will remain constant, and architectural/training changes will be made.\\
& Parameter tuning  & Observe training losses and make changes to training iterations, learning rates, and data batch sizes based on validation results. \\
& Quantitative feedback & Use image quality assessment metric to evaluate model's enhancement capability. \\ 
\hline
Specify scalability issues & Scaling the model & Change either the width or depth of the model to scale the DL model\\ \hline
Choose tools to use to evaluate the model &  \multicolumn{2}{l}{* Compute image quality assessment metric (PSNR / SSIM, WS-PSNR / WS-SSIM)} \\ \hline
Evaluate the quality of the model & \multicolumn{2}{l}{\begin{tabular}[c]{@{}l@{}}This is evaluated based on the number of parameters, runtime, and evaluation results \end{tabular}}\\ \hline
\end{tabular}
\end{table}

\begin{figure}[h!]
   \centering
\includegraphics[width=1\linewidth]{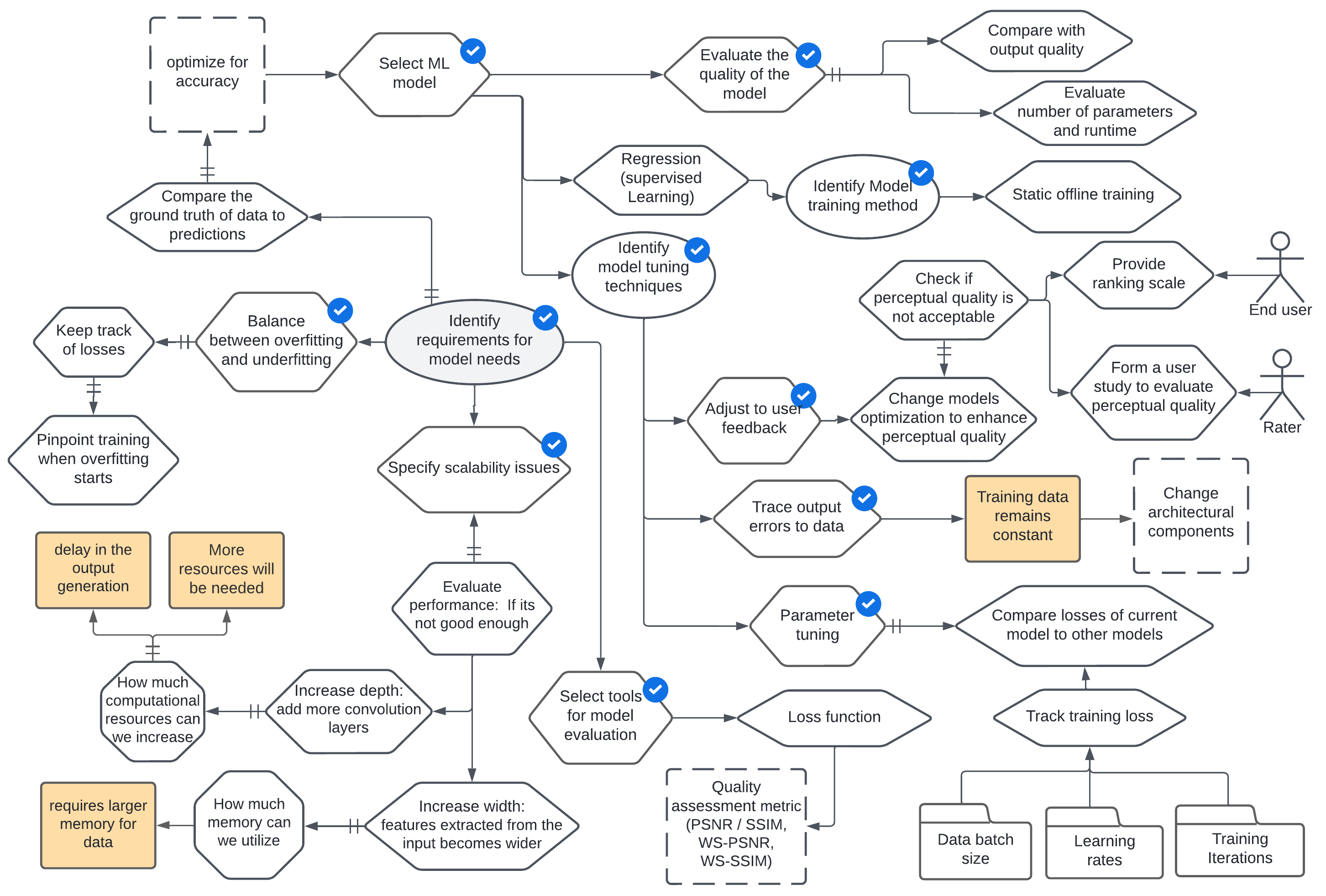}
\caption{Model presenting the requirements for Model Needs.}
\label{fig:ModelNeeds}
\end{figure}

Trade-offs that needed to be calculated and modeled were connected to scaling the LM. Scaling was achieved by changing the width or depth in the deep learning model~\cite{kawaguchi2019effect}. Increasing the depth meant adding more convolution layers and allowing the model to learn more complex patterns. The depth would have to increase if the performance were not good enough. However, the trade-off to increasing the depth is that more computational resources are required, resulting in delays in the output generation. On the contrary, changing the width would require that features extracted from the input become wider. This would require more memory allocation for data, as it would store more features for the given input at multiple layers. For the model used, an experiment would be set to identify when to modify width or depth and at what intervals while evaluating the trade-offs for time vs. costs to achieve the most effective results. Table~\ref{table:ModelNeeds} shows the collected requirements for model needs.

\begin{table}[h]
\caption{Identified data needs requirements for the 360 video enhancer.  * Requirements that could be specified before testing the data on the selected model}
\label{table:DataNeeds}
\centering
\footnotesize
\begin{tabular}{p{2.5cm} p{4cm} p{8.5cm}}
\hline
\multirow{7}{*}{Data selection} & \multirow{4}{*}{ Features} & Color: RGB \\ & & format: {360\textdegree} Videos \\  & & Representation: Equirectangular Projection \\ &  & size: 20 frames \\
 & Labels & N/A \\
 & Example & An actual video frame\\
 & Sampling rate & Up-to 20 frames from a given video clip \\ \hline
\multirow{3}{*}{Data collection} & Identify type of data needed & * Diverse contents to include different objects, motions and lightning \\
 & Amount of data needed & Initially, the quantity and diversity were not enough, so a conventional video dataset was used to initialize the model training \\
 &Diversity of data & To ensure diversity a complexity analysis was conducted \\ \hline
\multirow{2}{*}{Constraints} & cost, accuracy, quality, time? & Quality: Resolution of 480x360 (hardware and memory limitations) \\ & &Size: up to 20 frames (memory and model limitations) \\ \hline
\multirow{5}{*}{Data source} &  Type of data source & * Open-source: from research labs, YouTube, open-source platforms \\
 & Is data responsibly sourced? & *  All sources of the data collected are given credit and citations \\
 & Using feedback as data & *  N/A \\
 & Charges in obtaining data & *  Time taken to gather and clean the data needed \\
 & What measures are taken to ensure the data is up to date? & *  Data was recently collected (less than a month ago) \\ \hline
\multirow{2}{*}{Split Data}                                 & Training data (Model tuning) &  90\% randomly selected for training \\
 & Testing data & 10\% out of 590 randomly selected for testing \\ \hline
\multirow{5}{*}{Data quality} & Accuracy & *  Data is filtered to remove sudden changes in scene, motion, and light. Rolling and credit information are removed. Each video is divided into shots, and static shots are removed.        \\
 & Completeness & Completeness was insured in a technical way to show the diversity that data presents \\
 & Consistency & Data is processed to become the same length and resolution \\
 & Credibility & N/A \\
 & Currentness  & * Creating video shots that represent unique single scene; thereby ensuring that each clip/shot has current data belonging to a continuous period of time \\ \hline
\multirow{2}{*}{Data requirements} & Protect personal information & We are not collecting usage/user related information \\
 & Does data comply with privacy and law & *  Videos used for training and testing are not shared with the end users. The user will have their own choice of videos to use!   \\ \hline
\multirow{3}{*}{Fairness} & Identify biases: & * Model might not perform equally for videos with larger motion or luminance changes. \\
 & Missing features & More diverse scenes and objects are needed \\
 & Under or over representative data & Possibility of same scene/object represented across multiple videos \\ \hline
\end{tabular}
\end{table}

\subsection{Requirements for Data Needs}~\label{sec:reqDataNeeds}

\begin{figure}[h!]
   \centering
\includegraphics[width=\linewidth]{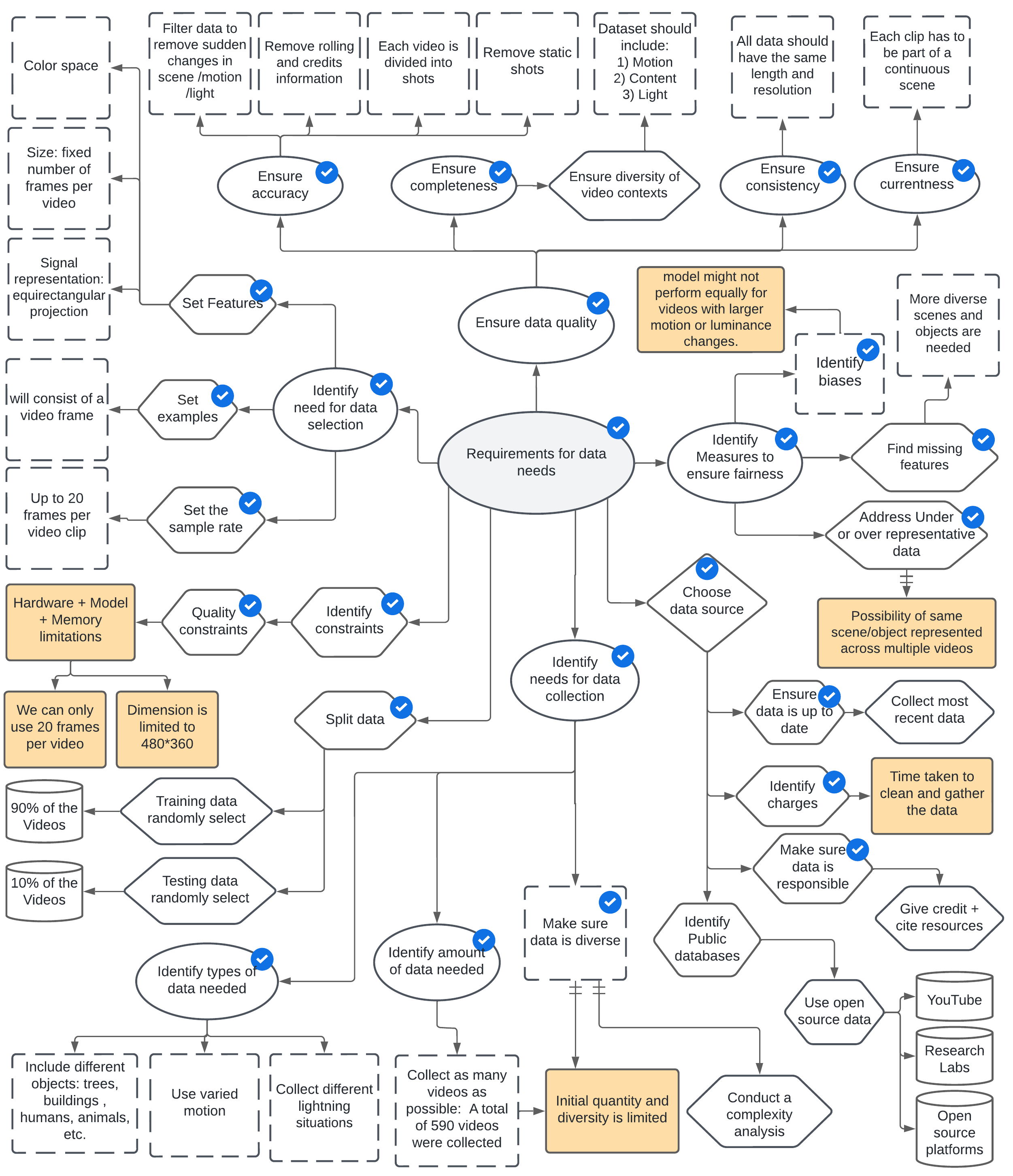}
\caption{Model presenting the requirements for Data Needs.}
\label{fig:DataNeeds}
\end{figure}

The third session involved extracting requirements for data needs. While conducting this session, it was found that the data collected originally was unsuitable for the model's training. Therefore, the data had to be cleaned and adjusted to become accurate, correct, consistent, and current. Sudden changes in scene, motion, or light and removing rolling and credit information were removed to make the data accurate. Next, to ensure data completeness, a list of all attributes and events needed for model training was listed, and the expert ensured each attribute had data to represent. Completeness was established technically to show the diversity of the presented data to ensure that the dataset had diversity in terms of 1) motion,  2) content, which included objects and scenes, and 3) light. The initially collected data was not consistent in quality. Thus, videos were processed to the same length and resolution to ensure consistency. Finally, currentness was provided by creating video shots representing a unique single scene, thus ensuring that each clip/shot had current data for a continuous period. 

Data requirements presented a higher number of needs. The need to specify how the data will be selected and collected for use in model training and testing. For data selection, the needs consisted of setting features, labels, examples, and sampling rates. Features include, among others, color, size, and location. An essential aspect of data needs was to identify any biases that might happen due to the data used. The proposed system was designed for videos with insignificant motion, content, and light changes across time. However, the model might perform differently if the videos had larger motion or luminance changes. Moreover, the diversity of the scenes and objects found was limited to what was available in the open-sourced platforms creating the possibility of the same scene/object represented across multiple videos. Thus, more diverse scenes would be needed to avoid under or over-representing data.  

Data collection involved gathering diverse content, including different objects such as trees, buildings, humans, and animals, and various motions such as how objects move across the video and lighting situations such as day, night, bright light, and dim. These measures were identified before data collection and obtained from literature based on the International Telecommunication Union's standard guidelines for spatio-temporal complexity measures~\cite{itu1999subjective}. Although measures were taken into account, some constraints were found after training the model with the collected dataset. The constraints were related to image quality and size due to hardware and model memory-related limitations. These constraints were modeled as limitations to show that data was limited to the specified size and frame rate.

\section{Discussion}~\label{sec:Discussion}

This section reflects on how applying our framework helped write and model requirements for AI-based systems. 

\subsection{RQ1. How does our RE4HCAI framework benefit the process of eliciting and modeling requirements for human-centered AI software?}

Most traditional RE methods are not equipped to manage AI-based software, and requirements are sometimes difficult to write, especially during the early stages of systems development. The problem with specifying requirements for an AI-based software project is that it is difficult to explain the black-box nature of AI and how the system would work or what outcome it will predict~\cite{belani2019requirements}. In our case study, we found that the process of eliciting and specifying requirements for AI-based software is vastly different from traditional approaches, as some aspects of the system are unpredictable and difficult to explain.

To evaluate the framework and modeling tool, we compared the building process of the {360\textdegree} VR model before and after the implementation of our framework. Before implementing the framework, many human-centered aspects were missing from the final product. Most of the development was based on the technical aspects of building the AI-based system. For example, classifying what data was needed was based on technical aspects (e.g., model and hardware needs) rather than focusing on the human perspective. After applying the framework, we identified possible biases that might lead to issues, e.g., biases related to the lack of diversity of data sources and the biases related to the lack of diversity of the type of videos.

The catalog helped elicit most of the requirements to build the AI-based software project from a human-centered perspective. The models presented in figures \ref{fig:UserNeeds}, \ref{fig:ModelNeeds}, and \ref{fig:DataNeeds} helped the ML expert visualize what needed to be considered when implementing the AI-based software. The conceptual models for our RE4HCAI framework provided a visual view of all the components in the system. Our participant said ``It helped me visualize all the components needed to be done in my project''. Also, we found that limitations and system needs were easier to spot with the visual representation via the conceptual model. Another observation we found missing was using feedback in the LM. The initial configuration of the LM did not consider the users' view of how good the perceptual quality of the improved {360\textdegree} videos was. However, after applying the framework, the decision was made to include user feedback to account for model tuning in future work of the {360\textdegree} VR project. 

We also noticed that some framework aspects would change depending on the application domain and the AI type used. For example, in this case study, regression was used, and the optimized reward function would be the loss function. Therefore, we need to consider other requirements that might need to be elicited and specified in our framework. This relates to Berry's finding~\cite{berry2022requirements} that different AI algorithms and their individual settings in different application contexts would require different requirements specifications. However, if the system were designed by applying a classification algorithm and we had to choose between precision or recall, one would need to calculate the trade-offs. When calculating the trade-off between recall vs. precision, we need to list the implications of having an FP vs. FN. In this case, if precision is favored, i.e., we would opt to reduce the number of FPs at the cost of tolerating more FNs, then the {360\textdegree} VR system will use the correctly selected TP frames but remove some frames from the outcome FN. Thus, the final outcome will have visual gaps, as some relevant frames will be missing. On the other hand, favoring recall meant that most frames would be included, i.e., more FPs. Thus, the final output will also include frames that are not enhanced FP, compromising the quality of experience as the visual quality will not always be the same in the final product, and some of the frames will still have the same quality as the original video.


The downside to favoring precision would be the possibility of having lags when rendering frames, as the system might choose to exclude frames that should have been enhanced from the frame sequence. These lags would increase the chance of latency which is a potential cause of VR sickness. Decreasing this delay or latency would result in reduced sickness \cite{mihelj2014virtual}. Recall would still include these frames when rendering, even if these frames might not have the expected quality. In this case, the trade-off is to compromise the visual quality instead of causing lags or delays in the visual display.

\subsection{RQ2. How can the RE practices be aligned in the life cycle of AI projects?}

When applying the framework to the case study, not all requirements could be elicited at the start of the project. We found that it was important to identify some requirements at the start. For example, when identifying user needs, we found that it was necessary to have initial specifications before building AI-based software. Also, some of the capabilities could be identified only after getting initial results from the model training, such as how much it would improve the quality of experience and if it would affect the quality of service. Furthermore, limitations such as hardware resources and processing time could not be identified at the start of the project. The extent of the limitation was found after training the model with the existing dataset. The reason was that the ML expert could not know how much scaling the model needed. Adding more features and layers resulted in higher processing time and required more resources to accommodate the change. They could establish how many features or layers were needed to improve model performance only after several training iterations. 

Similar patterns were found in both requirements for model needs and data needs. We observed that these requirements would change over time as they learned how the model would interact with the available data. For example, information regarding the needed frames could be specified only after initial testing. The datasets were limited to using up to 20 frames per clip, which was unknown at the beginning of the project. However, after initial testing, it was found that this was a constraint due to the model and hardware limitations. The same applied to identifying the quantity and diversity of the data. After experimenting on the first round of training and testing, they found that the existing 360 dataset was inadequate. Thus, the dataset had to be modified accordingly.

We established that it was important to specify requirements for data needs, such as data sources, charges, diversity of collected data, data quality, if data was up to date, and other requirements involving cleaning the data to provide accurate and current results. Some of these data requirements were obtained from literature and standard guidelines. Therefore, not all requirements could be specified at the start of the project, and some might change or appear while model testing is in progress. We highlight the requirements that we could identify prior to starting the project in tables \ref{table:UserNeeds}, \ref{table:ModelNeeds}, and \ref{table:DataNeeds} by adding a * character before the requirement.

\section{Threats to Validity}~\label{sec:Threats}

\textcolor{Black}{In all the phases of our research method and case study design, we attempted to mitigate and reduce any threat to validity as follows :}

\subsection{Internal Validity:}

Case studies are easier to perform than experiments., but more challenging to interpret~\cite{kitchenham1995case}. The main disadvantage to conducting case studies is that the outcomes are more susceptible to researcher bias, and evaluation usually depends on how \textcolor{blue}{the researchers interpret the results~\cite{easterbrook2008selecting}.} To reduce potential threats and biases, we ensured when selecting the case study that the person leading the AI project had no prior knowledge of our framework. Also, we collected requirements based on our framework and compared the results to how it was built without applying the proposed framework.

\subsection{Construct Validity:} 

\textcolor{Black}{Another threat was the selection of studies to build the catalog.} The risks of building a catalog for requirements are that it might miss out on some requirements and not be comprehensive~\cite{frank2013domain}. We established our requirements for the framework based on the guidelines and the literature presented on human-centered AI. Although we might have missed out on some requirements, we did base our findings on our SLR, which covered a comprehensive list of studies on RE4AI research. Also, we used industrial guidelines such as Google and Microsoft, which have already done extensive research to provide their guidelines. 

\subsection{External Validity:}
\textcolor{Black}{The framework has been applied to only one case study, which threatens external validity. We will address this threat by conducting more case studies in future work and investigating how the framework applies to projects from different application domains and multiple stages in the software development lifecycle. Also, due to time limitations, we could only assess the first three areas of the framework and plan to conduct a further evaluation to assess the last three areas in future work.}

\section{Related Work}~\label{sec:RelatedWork}

AI-based software should be carefully assessed not to replace people's abilities but augment their capabilities and allow people to make the final choice \cite{how2020artificial}. Shneiderman \cite{shneiderman2020Three} explained the importance of enabling humans to control the AI-based system. Human needs and values in building AI-based software need to be researched and examined carefully. For example, AI bots such as Alexia and Siri require extensive training to obtain the right personality \cite{wilson2018collaborative}, and some companies are investing in creating algorithms for chatbots that can detect sarcasm or respond with empathy~\cite{daugherty2018human}. 

Recent studies have shown that many AI-based systems lack requirements specifications~\cite{lwakatare2019taxonomy, kuwajima2020engineering,Zamani:21}, which is mainly due to the difference in the building process between traditional systems vs. AI-based software~\cite{sculley2015hidden}. We conducted an SLR \cite{ahmad2021s} to identify literature on RE4AI. From the results, we identified existing frameworks in RE4AI to include ~\cite{bosch2018takes, nalchigar2021modeling, aydemir2018roadmap, becker2020partial}. Nalchigar et al.~\cite{nalchigar2021modeling, nalchigar2018business} offers a framework to manage RE in building ML systems. The presented GR4ML framework covered three views: the business view, the analytic design view, and the data preparation view. A conceptual modeling language was used to present each view visually. The framework was applied to a case study that investigated the use of ML in the medical domain.     

Most AI-based systems are guided by large amounts of data and limited given resources. Bosch et al. \cite{bosch2018takes} proposed the Holistic DevOps framework for building AI-based software. The framework combined three practices: requirements-driven, data-driven, and AI-driven software systems. Requirements-driven approaches are used for systems that don't require frequent changes. This method was found to use fewer resources when testing the system \cite{tuncali2019requirements}. On the other hand, data-driven approaches are used for systems that are constantly changing and updated \cite{bosch2018takes} and are designed based on the analysis of available data \cite{bach2017data}. Companies that usually use automation, such as speech and image recognition, tend to use AI-driven approaches \cite{bosch2018takes}. 

Aydemir and Dalpiaz~\cite{aydemir2018roadmap} proposed a framework to aid requirements engineers and stakeholders in analyzing ethical requirements. The framework assisted in extracting, managing, and evaluating ethical requirements. However, the focus was on ethical requirements only and did not include other aspects of human-centered AI requirements. The last framework \cite{becker2020partial} provides a virtual framework to test specified requirements for a self-driving car. Also, frameworks are built to manage aspects of AI-bases systems, such as Khalajzadeh et al. \cite{khalajzadeh2019bidaml,khalajzadeh2020visual} who created a domain-specific modeling tool to support data analytic solutions for ML systems (BiDaML). The toolset provides five diagrams to support a different aspect of big data analytics. 

Shneiderman \cite{shneiderman2020bridging} presented a framework that proposes 15 recommendations for engineering trustworthy, reliable, and safe human-centered AI-based systems. To ensure reliable human-centered AI, software engineering teams need to apply technically sound practices such as using appropriate analysis tools, updating workflows for each task and domain, applying new forms of validation and verification, testing for bias detection, and providing explainable user interfaces. For reliable human-centered AI, there should be more commitment and training to ensure safety measures, reporting, and addressing errors and failures. And lastly, promoting trustworthy systems by providing independent oversight reviews to support legal and ethical codes of conduct. Another method used to aid in creating awareness of ethics when building AI software is the ECCOLA~\cite{vakkuri2021eccola} method, which is based on the ethical AI guidelines and involves using a deck of 21 cards, with each card targeting an aspect of ethics.

Although some frameworks are proposed to provide methods for managing RE4AI, none have focused on delivering an overall solution to building human-centered AI in RE. An exception is Shneiderman's framework~\cite{shneiderman2020bridging} that relies on promoting human-centered AI-based software. However, it focuses on the entire Software Engineering process, not RE. This calls for researching and studying appropriate human-centered AI methods before including them in software systems. We argue that the proposed solutions in RE should address human needs, as having a very well-engineered product would not be useful if it does not satisfy the user's needs.

\section{Conclusion \& Future Work}~\label{sec:Conclusion}

This paper offers a framework to extract and model requirements for human-centered AI. The proposed framework consists of three layers. The first layer provides a reference map to six areas from the human-centered AI guidelines. The second layer presents a catalog to elicit these requirements. The third layer provides a modeling tool to show the second layer's elicited requirements visually. The framework was applied to a case study, and we extracted requirements for the first three layers of user, model, and data needs. The case included specifying and modeling requirements for an AI system that enhanced the quality of 360{\textdegree} VR videos. We found that some requirements were more difficult to specify initially and could be identified only after testing the model with the existing data.

We plan further to evaluate our framework in the future in several workshops and compare it with existing frameworks and modeling platforms. Also, we do not provide details of how attributes are connected in sub-level models. For example, in model needs, when modeling the goal for balancing overfitting and underfitting, we did not model the actual process. This would have to be presented in a separate sub-model. We plan to extend our model to present this level of detail in future work.


\printcredits

\bibliographystyle{elsarticle-num}

\bibliography{refs}

\newpage
\section*{Appendix}
\small
\begin{spacing}{1.2}
\begin{longtable}{p{3cm} p{6cm} p{6cm}}

\caption{Catalog presenting requirements needed for Human-centered AI}
\label{table:Catalog}
\\\hline
\multicolumn{3}{l}{\begin{tabular}[c]{@{}l@{}}\textbf{User Needs}\end{tabular}} \\\hline \hline
\multirow{2}{9em}{Can AI add value or solve the problem to user’s needs?} & Do you have the expertise to manage AI? (Data scientist, ML specialist, SE) & IF no – use a non-AI solution \\
& Is data available? & (format – amount – diverse)   \\ \hline
\multirow{3}{*}{Identify need for AI?} & Who are the users &  \\
& Why do we need the system &  \\
& What is the system used for &  \\ \hline
\multirow{4}{9em}{What are the system capabilities?} & 
\multirow{2}{*}{Limitations} &  Limitations to the user:\\
 & & Limitations to stakeholders: \\
 & Capabilities (What can the system do) 
 & How well can the system do what it does?
 \\
 \hline
Interaction with user action & Proactive vs. Reactive & (Provides results when people request them) or Reactive (Interacts with the user without requesting) \\
\hline
Is the user aware of AI feature? & Visible or Invisible features &  \\
\hline
Evaluate approach & Augmentation vs Automation & (user enjoyment, control, high stakes) vs (boring, dangerous, cannot be performed by humans) \\
\hline
\multirow{8}{9em}{Reward function or evaluation matrix} 
 & Impact of having a false positive &  \\
 & Impact of having a false negative &  \\
 & Trade off - e.g precision vs recall &  Precision (excludes relevant results, removes all false positives, and misses some of the true positive predictions) vs recall (includes irrelevant results as it includes all true positive but might also include some false positive.) \\
 & Choice of reward function & e.g precision or recall \\
 & Impact on user &  \\
 & List potential pitfalls &  \\
 & How do you provide inclusion & \\
 & Predictions &(How are predictions used in making decisions)\\ \hline
\multirow{3}{*}{Predictions} 
 & How are predictions used in making decisions &  \\
 & Impact of new input on predictions &  \\
 & Impact of feedback on predictions &  \\ \hline \hline
 
 \multicolumn{3}{l}{\begin{tabular}[c]{@{}l@{}}\textbf{Model Needs} \end{tabular}} \\\hline \hline
 Optimize algorithm for? & Optimize for accuracy, explainability, robust, etc. &  \\ \hline
Choose the ML type & Supervised , Unsupervised, Reinforcement learning  &   \\ \hline
Model training & Static vs Dynamic & (Offline evaluation / Training improves with updates) vs (Learns from user behavior, Improves from feedback) \\ \hline
\multicolumn{2}{l}{\begin{tabular}[c]{@{}l@{}}Balance between overfitting \& underfitting\end{tabular}}  &   \\ \hline
\multirow{5}{10em}{Model Tuning (Include parameter tuning and architecture changes)}  & User feedback &  \\
 & Adjusting to user behavior & \\
 & Trace output errors to data & \\
& Parameter tuning  & \\
& Quantitative feedback &  \\ 
\hline
\multicolumn{2}{l}{\begin{tabular}[c]{@{}l@{}}Specify scalability issues \end{tabular}} & \\ \hline
\multicolumn{2}{l}{\begin{tabular}[c]{@{}l@{}}Choose tools to use to evaluate the model \end{tabular}} &   \\ \hline
\multicolumn{2}{l}{\begin{tabular}[c]{@{}l@{}} Evaluate the quality of the model \end{tabular}} &\\ \hline \hline


\multicolumn{3}{l}{\begin{tabular}[c]{@{}l@{}}\textbf{Data Needs} \end{tabular}} \\\hline \hline

\multirow{4}{*}{Data selection} &  Features & (a characteristics of an input variable.  E.g. Feature would include color, size, location, etc) \\
 & Labels & descriptions given to data (Explicit -manually vs Implicit – model learns) \\
 & Example & (a row of data and contains features and labels)\\
 & Sampling rate & amount of data in a given dataset \\ \hline
\multirow{3}{9em}{Data collection (Match data to user needs)} & Identify type of data needed \\
 & Amount of data needed &  \\
 & Diversity of data & \\ \hline
Constraints & cost, accuracy, quality, time? & \\  \hline
\multirow{3}{*}{Data source} &  Type of data source & (public, private, mixed)\\
 & Is data responsibly sourced? &  \\
 & Using feedback as data &  \\
 & Charges in obtaining data & \\
 & Measures taken to ensure data is up to date? &  \\ \hline
\multirow{2}{*}{Split Data}                                 & Training data (Model tuning) &   \\
 & Testing data & \\ \hline
\multirow{5}{*}{Data quality} & Accuracy & (The correctness of the data collected)    \\
 & Completeness & (all attributes and events should have data to associate them).    
-	List all attributes and events needed for the AI product
-	Make sure each attributes has data to represent
 \\
 & Consistency & (Collected data has to be free from contradictions)  \\
 & Credibility & (Credible data had to be authentic and truthful) \\
 & Currentness  & (corresponded to having data collected within the correct time frame; for example, when collecting images of people for a facial recognition application, a picture of an adult person when they were a baby will not be the correct data) \\ \hline
\multirow{2}{*}{Data requirements} & Protect personal information & \\
 & Does data comply with privacy and law &  \\ \hline
\multirow{4}{*}{Fairness} & Identify biases & \\
 & Missing features &  \\
 & Under or over representative data & \\ 
  & Other & \\ \hline
\multirow{5}{*}{Reporting biases} & Automation bias & Automation biases are when preferences are selected based on automated suggestions from the system\\
 & Selection bias & Selection bias usually happens when data is not collected randomly from the target population but rather selected based on the stakeholder's requests \\
 & Impact bias & Making an assumption based on the persons judgment (e.g. all wedding dresses are white)\\ \hline
 & Group attribution & Group attribution assumes that an output suitable for an individual will have the same impact on everyone in the group \\ 
  & Other & \\ \hline 
 Address biases & What methods will be used? & \\ \hline \hline


\multicolumn{3}{l}{\begin{tabular}[c]{@{}l@{}}\textbf{Feedback \& Control} \end{tabular}} \\\hline \hline
\multirow{6}{9em}{Explicit feedback (Only ask when needed)} & What reward will be provided to the end-user or raters to provide explicit feedback & (Material / symbolic) \\
 & If the reward is symbolic & What impact will it have on the user? What benefit will the user gain from it? \\
 & How is it going to be reviewed and evaluated? &\\ 
 & What changes will it make to the AI model? & \\
 & How will it be used in model tuning? & \\
 & How is feedback going to be conducted & (surveys, notifications, ratings, etc.) \\ \hline
 \multirow{3}{*}{Implicit feedback} & User interactions / behaviours to use: & Frequency of use, time of use, amount of time spent interacting with the system, accept / reject recommendations, etc.\\
  & Review feedback from user interaction to how it will make changes to AI &\\ 
 & How will it be used in model tuning & \\ \hline
\multirow{3}{*}{Calibration} & Reason for using calibration & \\
  & When should the user use calibration &\\ 
 & How is it going to affect the AI  & \\ \hline
\multirow{4}{9em}{When asking for feedback} & Measures taken to secure user privacy & \\
& Provide multiple options to feedback & (what options will you provide?) \\
  &  Is feedback in line with user mental map or users understanding of the system &\\ 
 & Allow the user to dismiss feedback or opt-out   & (How are you going to provide this option?) \\ \hline
 
 \multirow{4}{*}{User control} 
 & When to give the user control  & (High stakes, legal, safety )\\
  & Allow user to adjust preferences &\\ 
  & Allow user dismissal & \\
 & Level of control   & (high, medium, low or no control) \\ \hline \hline
 
    
\multicolumn{3}{l}{\begin{tabular}[c]{@{}l@{}}\textbf{Explainability \& Trust} \end{tabular}} \\\hline \hline \multirow{2}{*}{Explain to stakeholders} 
  & Limitations of AI & \\
  & Capabilities of AI &\\  \hline 
 \multirow{2}{*}{Explain to user} 
 & Inform user when changes happen & \\
  & Explain consequences to users action &\\ \hline
 \multirow{3}{*}{Explain data:} & Who can access or use the user’s personal data & \\
  & How data is shared between apps &\\ 
 &  Explain how predictions are based on data & \\ \hline
\multirow{3}{*}{Explain predictions} 
 & Explain with examples & \\
  & Explain with confidence: & Data visualization (Expert users).  Numerical (Might cause confusion).  Categorial (e.g., Low - Med - High).  N-best Alternatives (Low confidence) \\ 
 & Do not display Confidence:  & Risks (misleading, no impact to user decisions when explaining output, distracting).  Low confidence (Can provide partial explanation
\\ \hline
 Calculate Trade-off &	Identify conflicts with other requirements & \\ \hline
 \multicolumn{3}{l}{\begin{tabular}[c]{@{}l@{}}\textbf{Calibrate user trust} \end{tabular}} \\\hline
 \multirow{2}{*}{Explain special cases}
 & Law \& rules & \\
  & Third party involvement &\\  \hline
 \multirow{2}{*}{Explain feedback}
 &Impact of feedback on AI & \\
  &  How feedback is used to improve model &\\ \hline
 \multirow{3}{*}{Avoid over trusting}
&  Only show relevant info & \\
& Account for different situations  & -	Low stakes.  High stakes\\
& Explain errors & \\ \hline
   
\multicolumn{3}{l}{\begin{tabular}[c]{@{}l@{}}\textbf{Errors \& Failures} \end{tabular}} \\\hline \hline

\multirow{3}{9em}{User perceived errors (might change over time)}
& Failstates & True negative (Not included in system training).  System limitation \\
  & Context errors (these are true positive but) &	AI making incorrect assumptions.  User has poor mental model.  System not aligned with users needs  \\ \hline
Background errors & Invisible errors the user cannot perceive & \\ \hline
\multirow{13}{9em}{List possible errors source }
& \textbf{Prediction \& data errors} \\
  & Mislabeled or misclassified results &	Poor training.  Incorrect labeling \\
& Incorrect Model	& \\
& Incomplete data	& \\ 
& \textbf{Input Errors} & \\
& Unexpected input 	& \\
& Old habits	& \\
& Mis-calibrated input	& \\
& \textbf{Output Quality} & \\
& Low confidence &	Lack of data. Uncertain prediction accuracy. Unstable information \\ 
& \textbf{systems hierarchy error} & \\
& Multiple systems &	(Allow user to give one of the system priorities) \\
& Crashing signals &	(Allow signals from the primary system only) \\\hline
 
 \multirow{2}{*}{Error risks}
& High stakes & Health, safety and financial decisions
-	Sensitive / private data \\
  & Abusive users &  \\ \hline
 
 \multirow{3}{*}{Action}
& Mitigate Errors & \\
  & Allow users to fix mistakes &\\ 
 & Provide suggestions when in doubt  & \\ \hline
\end{longtable}
\end{spacing}

\end{document}